\documentclass[11pt]{article}

\usepackage{hyperref}

\usepackage{graphicx}
\usepackage{amsmath,amssymb}
\usepackage[utf8x]{inputenc}
\usepackage{cite}
\usepackage{nameref,hyperref}
\usepackage[table]{xcolor}
\usepackage{array}

\begin{document}
\vspace*{0.2in}

\begin{flushleft}
{\Large
\textbf\newline{\huge \bf Evolution of default genetic control mechanisms}}
\newline
\\
William Bains\textsuperscript{1}*,
Enrico Borriello\textsuperscript{2},
Dirk Schulze-Makuch\textsuperscript{3,4,5}
\\
\bigskip
\textbf{1} Department of Earth, Atmospheric and Planetary Sciences, Massachusetts Institute of Technology. Cambridge. MA 02139, USA 
\\
\textbf{2} ASU-SFI Center for Biosocial Complex Systems, Arizona State University, Tempe, AZ 85281-2701 , USA
\\
\textbf{3} Zentrum für Astronomie und Astrophysik, Technische Universität Berlin, 	Straße des 17. Juni 135, 10623 Berlin , Germany
\\
\textbf{4} German Research Centre for Geosciences (GFZ), Section Geomicrobiology, Potsdam, Germany.
\\
\textbf{5} Department of Experimental Limnology, Leibniz-Institute of Freshwater Ecology and Inland Fisheries (IGB), Stechlin, Germany.
\\
\bigskip

* Correspondence to W.Bains; bains@mit.edu

\end{flushleft}

\section*{Abstract}
We present a model of the evolution of control systems in a genome under environmental constraints. The model conceptually follows the Jacob and Monod model of gene control. Genes contain control elements which respond to the internal state of the cell as well as the environment to control expression of a coding region. Control and coding regions evolve to maximize a fitness function between expressed coding sequences and the environment. 118 runs of the model  run to an average of $1.4\times 10^6$ `generations' each with a range of starting parameters probed the conditions under which genomes evolved a `default style' of control. Unexpectedly, the control logic that evolved was not significantly correlated to the complexity of the environment. Genetic logic was strongly correlated with genome complexity and with the fraction of genes active in the cell at any one time. More complex genomes correlated with the evolution of genetic controls in which genes were active (`default on'), and a low fraction of genes being expressed correlated with a genetic logic in which genes were biased to being inactive unless positively activated (`default off' logic). We discuss how this might relate to the evolution of the complex eukaryotic genome, which operates in a `default off' mode.  

\section{Introduction}
Obligate multicellularity is uniquely a eukaryotic trait (1-3), and with it the morphological complexity that comes from combining many distinct cell types into one organism. Multicellularity requires complex genetic controls both to provide the control to generate different genetic activity patterns in different cell types and to provide the `programme' to construct the adult organism. In addition, the more complex internal architecture and controls in the eukaryotic cell also require specific controls. In some single-celled eukaryotes such internal complexity resembles that of equivalently sized multicellular organisms. Reflecting this, genome sizes in eukaryotes can exceed those of the largest bacterial or archaeal (``prokaryotic'') genomes by 4 orders of magnitude (Figure 1).

\begin{figure}[!b]
\centering
\includegraphics[width = .65\textwidth]{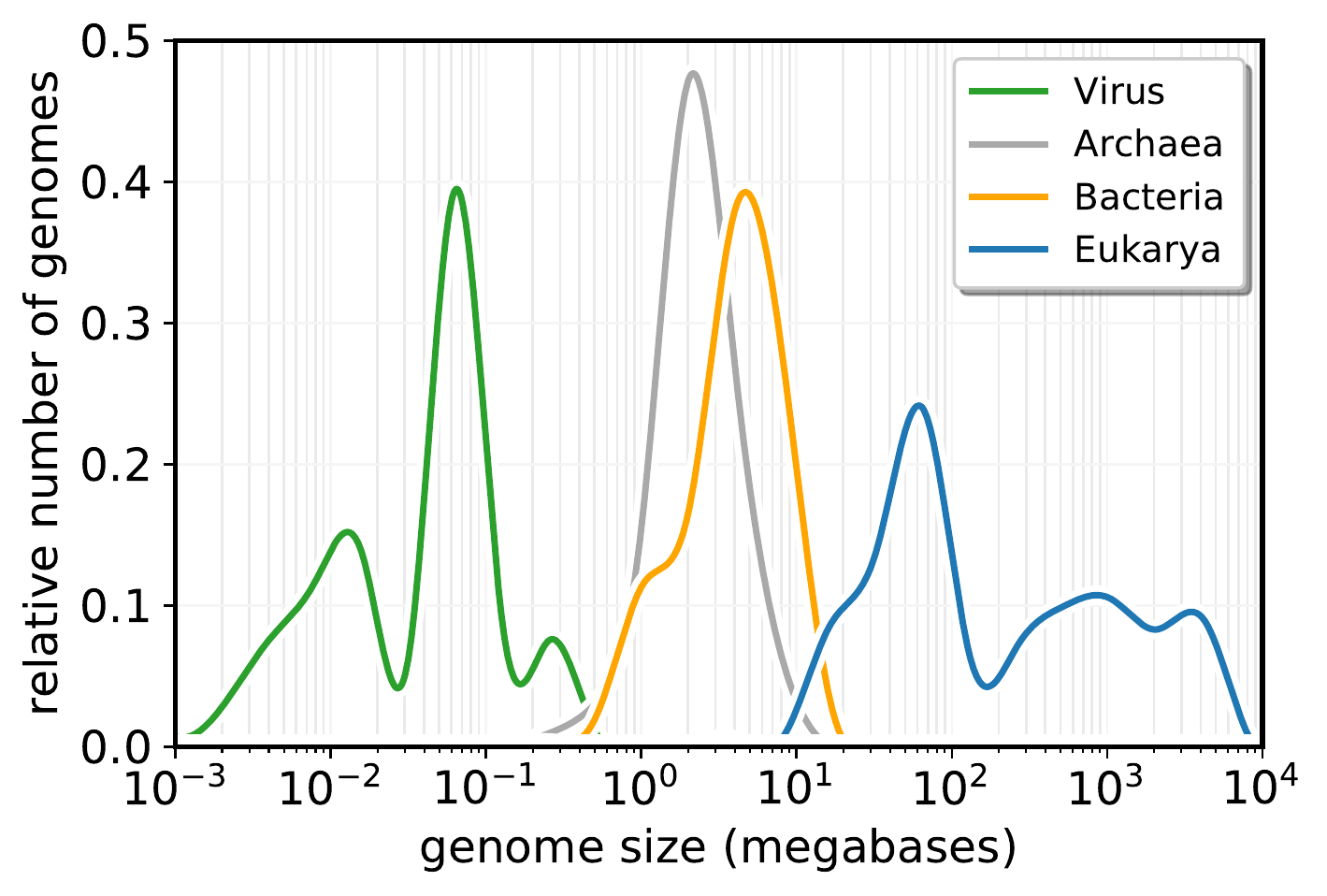}
\caption{Sizes of completed genome sequences, showing distinct sizes for prokaryotes (bacteria and archaea), eukaryotes and viruses. $X$ axis: genome size in megabases. $Y$ axis: fraction of each of the four classes of organism that have genome of that size. Data from \href{https://www.ncbi.nlm.nih.gov/genome/browse/\#!/overview/}{https://www.ncbi.nlm.nih.gov/genome/browse/\#!/overview/}, accessed 15th June 2020; based on 27308 bacteria, 1769 Archaea, 5300 Eukarya and 19536 viruses. }
\end{figure}

There is substantial overlap in coding capacity between the larger prokaryotic genomes and eukaryotic genomes. The coding capacity of some of the larger prokaryotic genomes such as those of some  cyanobacteria ($\sim12'000$ coding sequences (CDS) (4)), {\it Ktedonobacter racemifer} ($\sim11'500$ CDS (5)), {\it Sorangium cellulosum} ($\sim9000$ CDS (6)), {\it Magnetobacterium bavaricum} ($\sim8500$ CDS (7)) overlaps with coding capacity of multicellular fungi ($5000-15'000$ ( e.g. (8, 9)) and autotrophic protists ($10'000-20'000$ CDS (9)) and approaches that of Drosophila melanogaster ($\sim16'000$ CDS (10)). The size difference between prokaryotic and eukaryotic genomes is primarily due to non-coding DNA that is related in part to gene control. Thus the E.coli genome has little non-coding DNA, and $\sim285$ proteins are involved in gene control (11), $\sim7\%$ of the genome. By contrast over 90\% of the human genome is non-coding, and conservative estimates are that 10 times as many non-coding bases as coding bases are evolutionarily conserved (i.e. are presumed to have selectable function) (12, 13).  

What enabled this increase in genetic complexity? The key difference between prokaryotic and eukaryotic cells have been suggested to be chemistry, intracellular structure, energetics and genetics. In general, any small molecule structure made by a eukaryotic cell will be made by a prokaryote as well. Many  ‘eukaryotic’ cellular structures are actually found in a few prokaryotes as well.  Linear chromosomes are found in bacteria ($14-17$). Intracellular membrane compartments for secretion and processing (18, 19) and energy capture (19-24) as well as membrane-bound DNA-containing bodies are found in Planctomycetes (22, 25). {\it Achromatium oxaliferum} contains complex internal membranes containing calcium carbonate (whose function is obscure) (26),  {\it Entotheonella} detoxifies arsenic and barium by sequestering it in internal vesicles (27), and cyanobacteria have stacked internal photosynthetic membranes (28). The intracellular membrane system of eukaryotes is integrated into a dynamic network of vesicle trafficking and control which is rare in prokaryotes (reviewed in (29)); however some of the core proteins and structural elements of a cytoskeleton are also found in prokaryotes  ($30-35$), and the giant bacterium {\it Epulopiscium fishelsoni}  has an internal tubule system so similar to eukaryotes that it was initially mistaken for a protozoan (36, 37). These examples all suggest that complex structure per se follows from large size, rather than large size following from internal structure. 

It is widely accepted that the modern eukaryotic cell evolved by a series of endosymbiotic events (38, 39). Recent insights gained from molecular  biology show examples of endosymbiotic bacteria that live inside other bacteria ($40-43$), and bacteria that live inside modern mitochondria (44) (as well as a wealth of endosymbiotic bacteria in eukaryotic cells) which suggests that prokaryotic endosymbiotic events, while unusual, are not extremely rare. Lane and Martin ($45-47$) suggest that the endosymbiotic event, and consequent development of internal membrane-bound energy-generating organelles, enabled the ability to generate energy from intracellular membranes acquired through endosymbiosis is key, as more genes imply more proteins and proteins require energy to make. We find this theory lacking for three reasons. Firstly, the majority of genes in the larger eukaryotic genomes do not code protein – complexity comes from non-coding RNA genes and regulatory elements as discussed above. Secondly, most of the coding genes in any one cell are not transcribed; indeed the whole reason to maintain a complex genetic apparatus is so that different subsets of genes can be expressed at different times. Genomes containing more coding sequences do not make more proteins at any one time.  Lastly, protein synthesis is only the major use of cellular energy in autotrophic bacteria grown under conditions of unlimited nutrition. Under more normal conditions of growth, protein synthesis rarely is observed to consume more than 20\% of the cell’s energy, and of course in non-growing cells (which is most cells in the biosphere most of the time) protein synthesis is only needed for maintenance and turnover, a minor part of the overall ‘maintenance energy’ ($48-50$). (See Supplementary File 1 for a more detailed analysis protein synthesis’ energy requirements).  

We have recently suggested that the default logic of gene control is a significant factor enabling eukaryogenesis (51). It is observed  that it is easier for a gene to be expressed in a prokaryote than a eukaryote, as evidenced by the fate of pseudogenes, the construction of expression vectors, and the fate of differentiated gene expression in cell fusion (reviewed in (52)) as well as arguments from the mechanisms of gene control (See below). While there are exceptions, it is broadly true that the eukaryotic genome is by default ‘off’ and needs metabolic energy to turn ‘on’, whereas in a prokaryotic genome genes are by default ‘on’ unless turned ‘off’. 

This default control mode is reflected in the thermodynamics of gene control.  In eukaryotes, complexes of proteins are required to remodel chromatin around promoters before genes can be transcribed, involving the ATPase molecular motors Snf2 and Sth1 (53, 54), and subsequent ATP-dependent binding of transcription factors to chromatin  (55) before RNA polymerase can bind to a promoter. The remodelling involves (inter alia) ATP-dependent removal of H2A/B dimers from nucleosomes (56-59) (Archaeal nucleosomes lack H2A/B dimers, and consist of homologues of H3/4 dimers only (60, 61)). By contrast the molecular rearrangements that control initiation of bacterial transcription are powered by the binding energy of the various proteins (11, 62, 63). Archaea have similar transcription initiation logic to bacteria, despite having RNA polymerase complexes similar to those in eukaryotes (64, 65). RNA elongation in bacteria requires roughly 1.5 ATP per base added, again being controlled by protein binding factors (11). In Eukaryotes ATP-dependent chromatin remodelling is required for RNA elongation, as well as energy-consuming histone acetylation and methylation chemistry (66).  

We argue that this ‘Default off’ logic is more efficient if the majority of the genome is silent, as would be the case if the genome encoded many expression programmes only one of which is active at once. It would also allow the facile accumulation of silenced duplicate genes to act as the substrate for genome complexification, which itself is associated with rapid diversification and adaptation (67). We propose that a ‘default off’ logic will favour the evolution of complex genomes which code for multiple expression patterns, a ‘default on’ logic will favour the evolution of compact, efficient genomes with relatively few distinct phenotypes.  

This hypothesis should be testable by simulation and by experiment. As a first step in this we present a simplified model of gene control and evolution that can evolve either ‘Default On’ or ‘Default Off’ logic. In this paper we present the model, and initial results from its execution.

\begin{figure}
    \centering
    A)\hfill\phantom{0}\\
    \includegraphics[width=.75\textwidth]{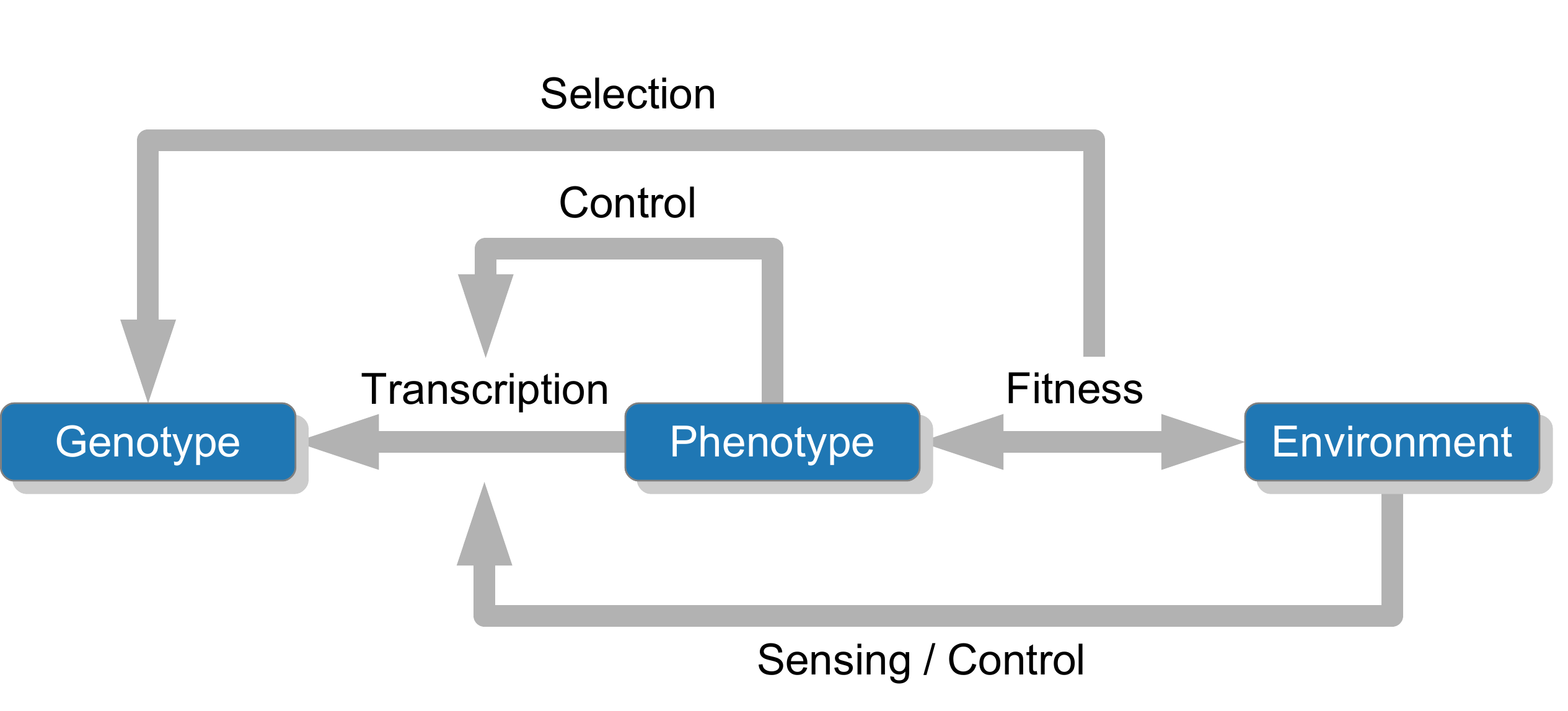}\\
    B)\hfill\phantom{0}\\
    \includegraphics[width=.75\textwidth]{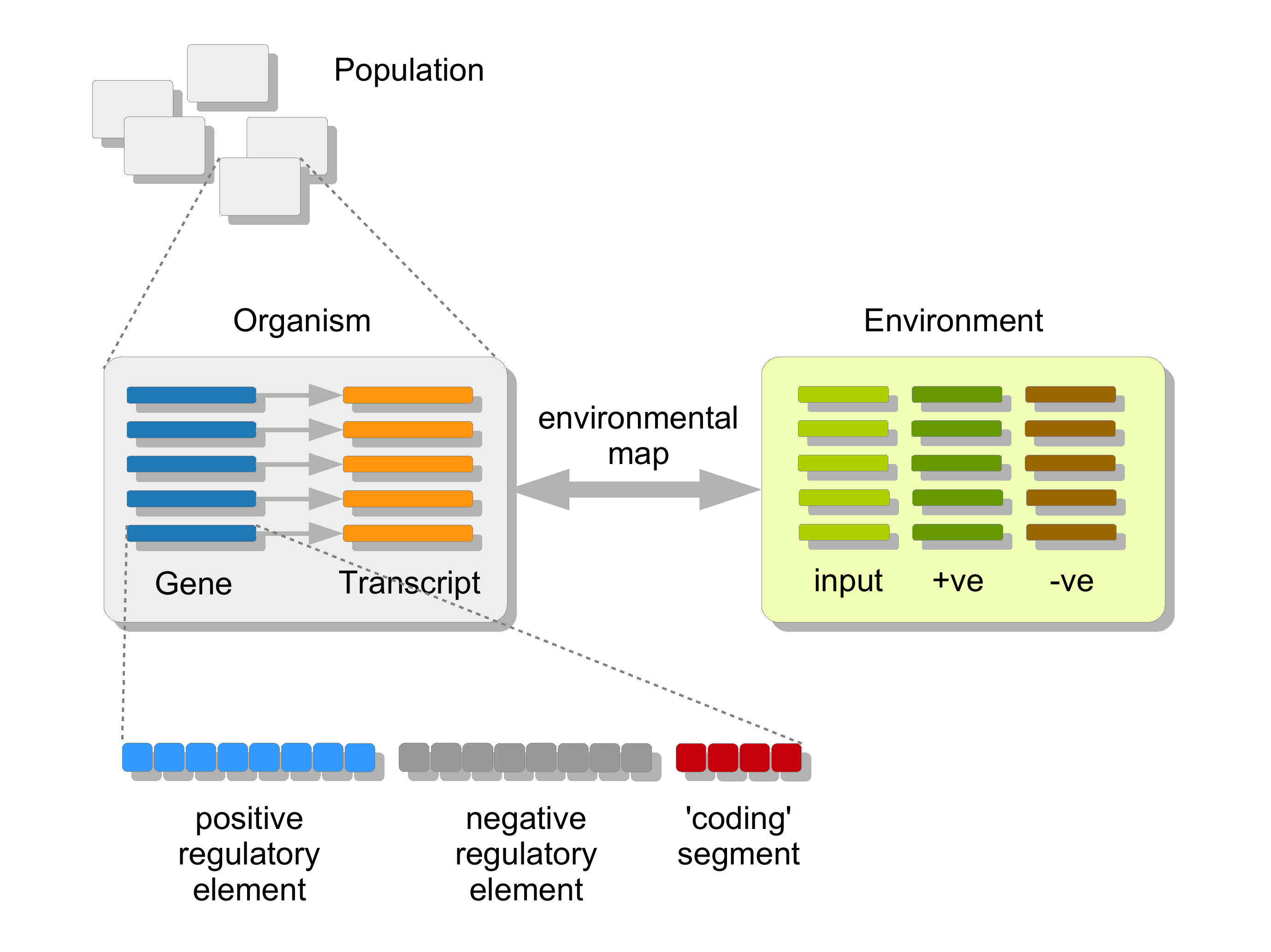}
    \caption{Summary of model structure. A) overall design philosophy, showing feedbacks between genotype, its encoded phenotype, and the environment that it must fit. B) summary of model components. See text for details. }
    \label{fig:my_label}
\end{figure}

\section{Methods}

    \subsection{Modelling approach}

We attempt to model the evolution of control logic of genes under selective pressure. As a balance between the need for computability on one hand and the need for biological ‘realism’ on the other, we chose the ‘classical’ operon as a model on which to build the model structure. A series of sequences upstream of the coding sequence can bind proteins which allow, promote or catalyse transcription (positive elements) or which can bind proteins that retard or prevent transcription (negative elements).  A similar process applies to eukaryotic genes in that positive and negative regulatory elements influence the transcription of the gene, although in eukaryotes those regulatory elements may be distant from the gene. Whether those regulatory elements are active will depend on the proteins in the cell, so that there is feedback between the phenotype and the transcription of the genotype that it encodes. The fitness of an organism depends on the ‘fit’ between its phenotype and its environment, but that environment can change, so the expression of genes must also be influenced by the environment. The model must also be able to be queried for some surrogate of ‘default off’ or ‘default on’ genetics {\it independent} of how many genes in an organism are actually transcribed at any one time (which will depend on the demands of the environment).  

The properties of the model are summarised in Figure 2A.

    \subsection{Specifics of the model}

To capture the requirements above, the model was constructed as follows. For simplicity, everything in the model is strings of one type. Thus the {\it phenotype} is a set of strings of the same sort of as the {\it genotype}. The strings are made up of different {\it characters}; there can be any number of types of characters (if the strings were to mimic DNA or RNA, the number of character types would be 4; the model was run with the number of character types ranging from 2 to 16). There is no equivalent of protein translation in the system. The model consisted of a number of {\it organisms} – in this initial implementation there were only 5 organisms for computational reasons. The organisms exist in an {\it environment}. Each organism contains a number of {\it genes}  which together comprise its {\it genotype}; in the runs reported here, organisms contained 25, 50 or 100 genes. Each gene is composed of up to ten {\it positive regulatory} elements, up to ten {\it negative regulatory} elements, and a {\it coding sequence}. The sum of the coding regions of genes that are active at any one time comprise the organism’s phenotype. The organism’s fitness is the match between its phenotype and its environment as follows. The environment comprises {\it positive elements}, {\it negative elements}, and {\it signalling elements}. Fitness is the sum of the number of positive environmental elements that match the current phenotype minus the number of negative environmental elements that match the current phenotype. (This is to reflect that sometimes having a function in a cell can be detrimental to the cell; were this not true, in our model all cells would express all genes all the time for maximum fitness.)  

Gene expression is controlled as follows. A regulatory element is active when either it matches an environmental signalling element or it matches the phenotype. This represents the transduction of an environmental signal into gene activity, and the transduction of internal gene activity into gene activity. If the sum of the number of active positive regulatory elements exceed the number of active negative regulatory elements then the gene it transcribed and its coding sequence is added to the phenotype. 

The model is seeded with random strings. At each cycle a new phenotype is computed, and new fitness computed for each organism , and the most fit organism randomly replaces one of the other organisms (which can include self-replacement). The organisms are then mutated by making small, random changes (character changes, insertions or deletions, with a bias of 6:4 deletion over insertion) to a fraction (typically between $10^{-5}$ and $5\times 10^{-5}$) of the strings in the genotype, or completely deleting one of them (typically with a probability between $10^{-6}$ and $5\times10^{-6}$).  

The model components are summarised in Figure 2B.

\section{Results}

    \subsection{Modelling selection and adaptation} 

We begin by showing that the model produces results that are consistent with adaptation, i.e. with changing from an initial random state to a state where the average fitness of the organisms is greater than it was at the start. We emphasise that changes made to the components of the model are entirely random; there is no directionality in the model except a slight bias towards gene shrinkage noted below. Both the initial genome and the environmental factors that the genome has to adapt to are randomly generated as well. Adaptation is therefore the result of selection for better ‘fitness’.  

We can measure the ‘degree of perfection’ $P$ of an organism in terms of the evolved fitness $F$ as a fraction of the possible maximum fitness, as the maximum fitness is the number of environmental factors $E_f$. Some example fitness curves are shown in Figure 3.  Figure 3A shows a typical curve that reaches a plateau of fitness and then does not achieve any greater fitness in the run.  Figure 3B shows a curve that is similar to $\sim500'000$ generations, but then a new increase in fitness is observed.  Figure 3C shows the decomposition of the fitnesses to each of three environments in a model, together with the average across all environments. In this run, the organism is tested against one of three, unrelated environments; the environment that the organism has to match changes every two generations. Note that fitness to each environment does not increase in parallel – sometimes selection has resulted in better fitness for one environment, sometimes for another. Fitness for an environment can actually decline if overall fitness does not decrease substantially. Figure 3D shows the separate fitness trajectories of five organisms as they evolve in a single environment. Again, individual organism can lose fitness, but the population trend is usually to increasing fitness. Lastly, Figure 3E shows a model that has not evolved significantly. Most of the change in fitness in Figure 3E appear to be noise, and fitness wanders around a low average ($P\sim0.04$ in this case, as $E_f=100$).  

\begin{figure}
    \centering
\begin{tabular}{cc}
\hspace{-2cm}A) \hfill \phantom{0}& \hspace{-2cm}B) \hfill \phantom{0}\\
  \includegraphics[width=.45\textwidth]{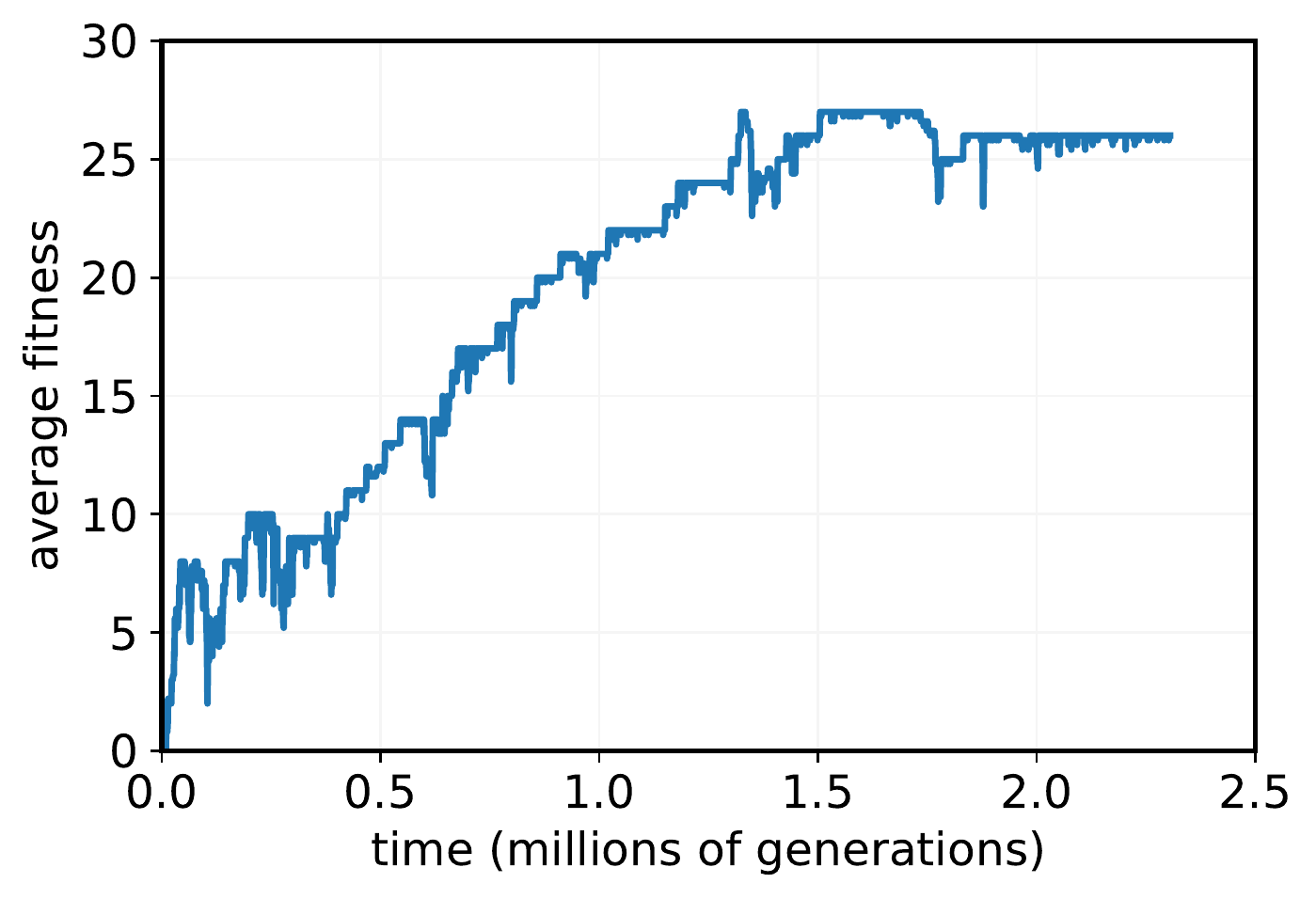}   &  \includegraphics[width=.45\textwidth]{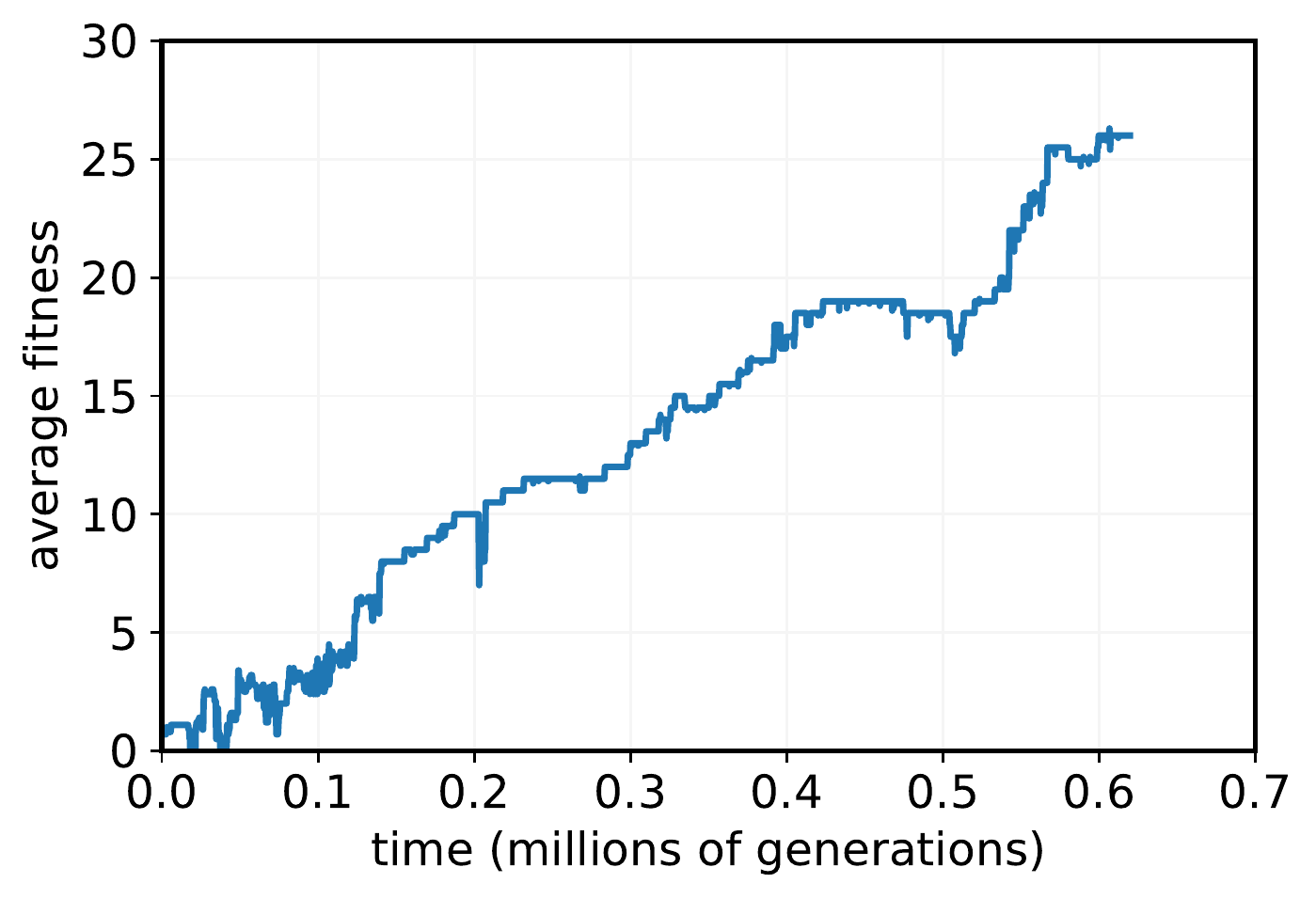}\\
  \hspace{-2cm}C) \hfill \phantom{0} &  \hspace{-2cm}D) \hfill \phantom{0} \\
  \includegraphics[width=.45\textwidth]{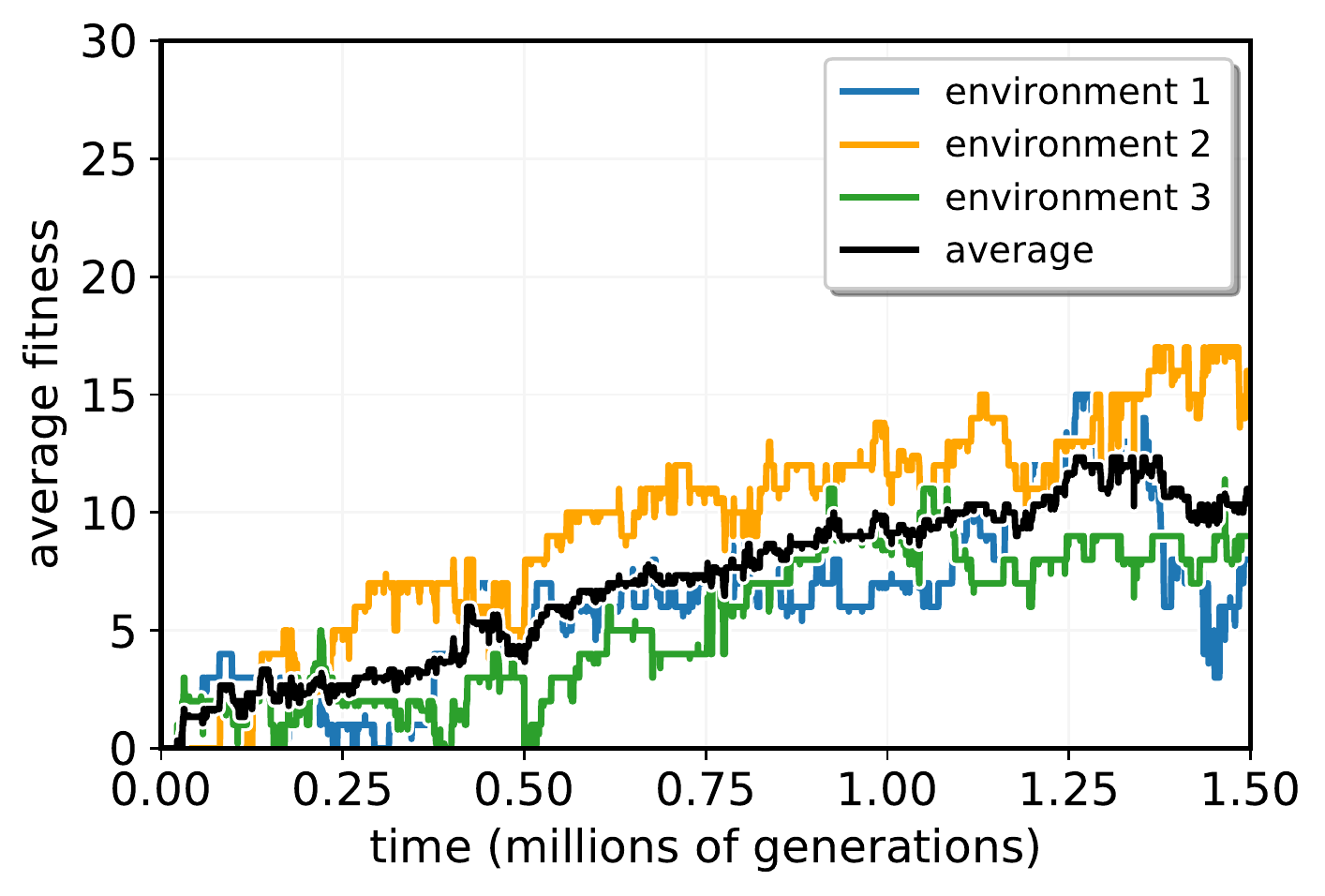}   &  \includegraphics[width=.45\textwidth]{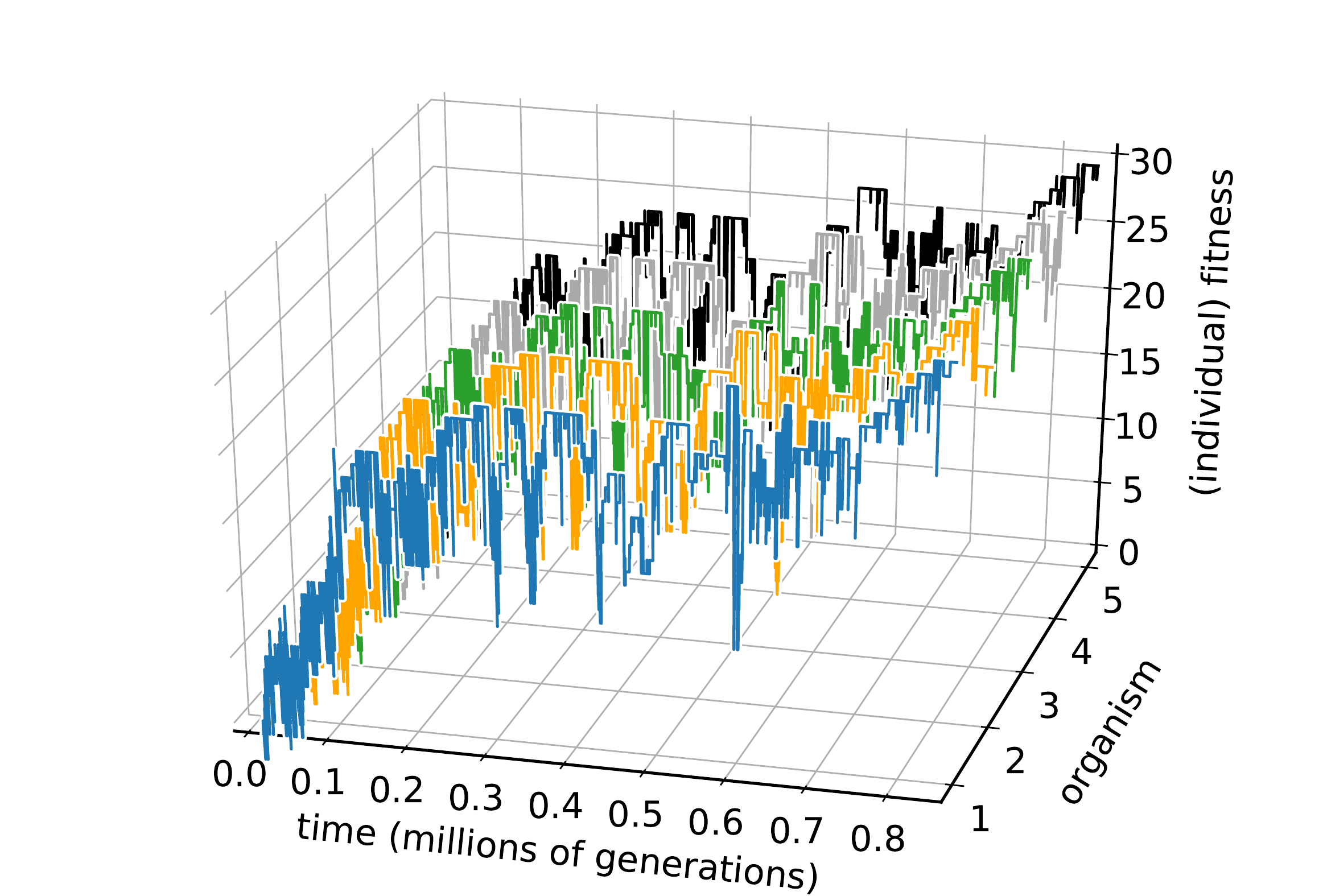}\\
  \hspace{-2cm}E)  \hfill \phantom{0} & \\
    \includegraphics[width=.45\textwidth]{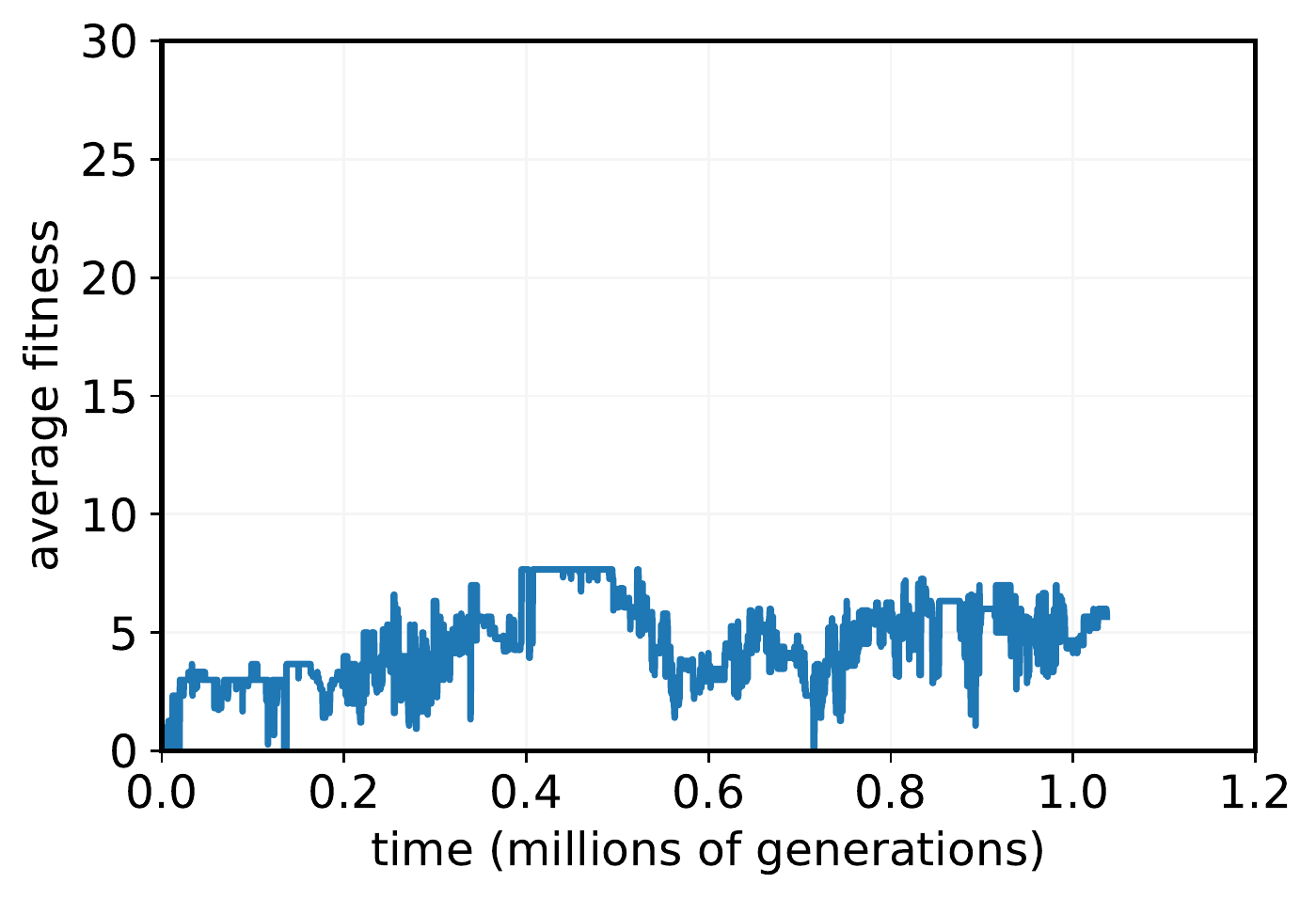}   &  \\
\end{tabular}   
    \caption{Examples of fitness plots for different runs of the model. A) Average fitness of the population converges smoothly on a maximum. B) Average fitness shows a jump in adaptation at $500'000$ generations. C) convergence of average fitness across three environments, showing divergent adaptation to each of the environments. D) Fitness of each of the five organisms making up the population plotted separately in a run that converges on a solution. E) Plot of fitness in a run that fails to converge on an optimum fitness.}
\end{figure}
    
    \subsection{Failure to adapt}

Models did not converge onto a fit state $\sim 1/5$ of the time (depending on what ‘fit’ means, and the selection of parameters).  This was found to be a function of the degree to which the genome complexity can match the environmental complexity (Figure 4). Highly complex environments are not efficiently matched by low complexity genomes, as would be expected. This to a degree is a consequence of limited run-time: some model runs had low fitness for a time and then experienced a ‘jump’ in fitness as a low-probability solution was ‘discovered’ (e.g. Figure 3B).  

We define whether a population is converging on a solution with a Curve Parameter $C_p$ as follows: we define a time $P$ as the time at which the population reaches a plateau of adaptation, i.e. does not appear by inspection to be able to increase its adaptation. $C_p$ distinguishes between populations that smoothly approach such a fitness plateau, such as shown in Figure 3A, and populations whose fitness fluctuates, such as in Figure 3E. Thus, if the fitness at times $0.25 P$, $0.5 P$, $0.75 P$ and $P$ are $A$, $B$, $C$, $D$ respectively, and $S(x)$ is the sign of $x$, (such that $x > 0 \Rightarrow s(x) = 1$;  $X < 0 \Rightarrow s(x) = -1$;  $x = 0 \Rightarrow s(x) = 0$) then  

\[
C_p = s (B-A) + s (C-A) + s (D-A) + s (C-B) + s(D-B) + s (D-C).  
\]

If $A < B < C < D$ (i.e. fitness is increasing throughout the run), then $C_p = 6$. 

For this analysis, runs of the model were only used if the curve parameter $C_p$ is greater than zero. 101 out of 118 runs of the model met this criterion. Omitting curves for which $C_p \leq 0$ resulted in substantially less scatter in the adaptation curves averaged across all models (Figure 5). 

Would all models converge on an optimal solution eventually? We hypothesise they would, but it might take years to achieve this using the relatively inefficient coding, which for practical reasons was run for only an average of $1.4 \times 10^6$ generations. (For comparison, the long-term evolution experiments performed by the Lenski lab. have been running for more than $20'000$ generations, and show a range of adaptations in gene control  structure without changing the underlying mechanisms or logic of the gene control architecture  ($68-70$)). Models were therefore stopped when they seemed to have reached a steady state of control structure (as defined below) and fitness within this time window. Future work will seek a more objective measure of termination state through spectral analysis of the fitness functions in an exhaustive scan of our parameter space, Disentangling the typical timescale of purely stochastic fluctuations from the timescale of selection-induced changes will provide a useful estimate of the average, expected time until convergence, as well as estimates of the range on that time.  

\begin{figure}[!h]
    \centering
    \includegraphics[width = .65\textwidth]{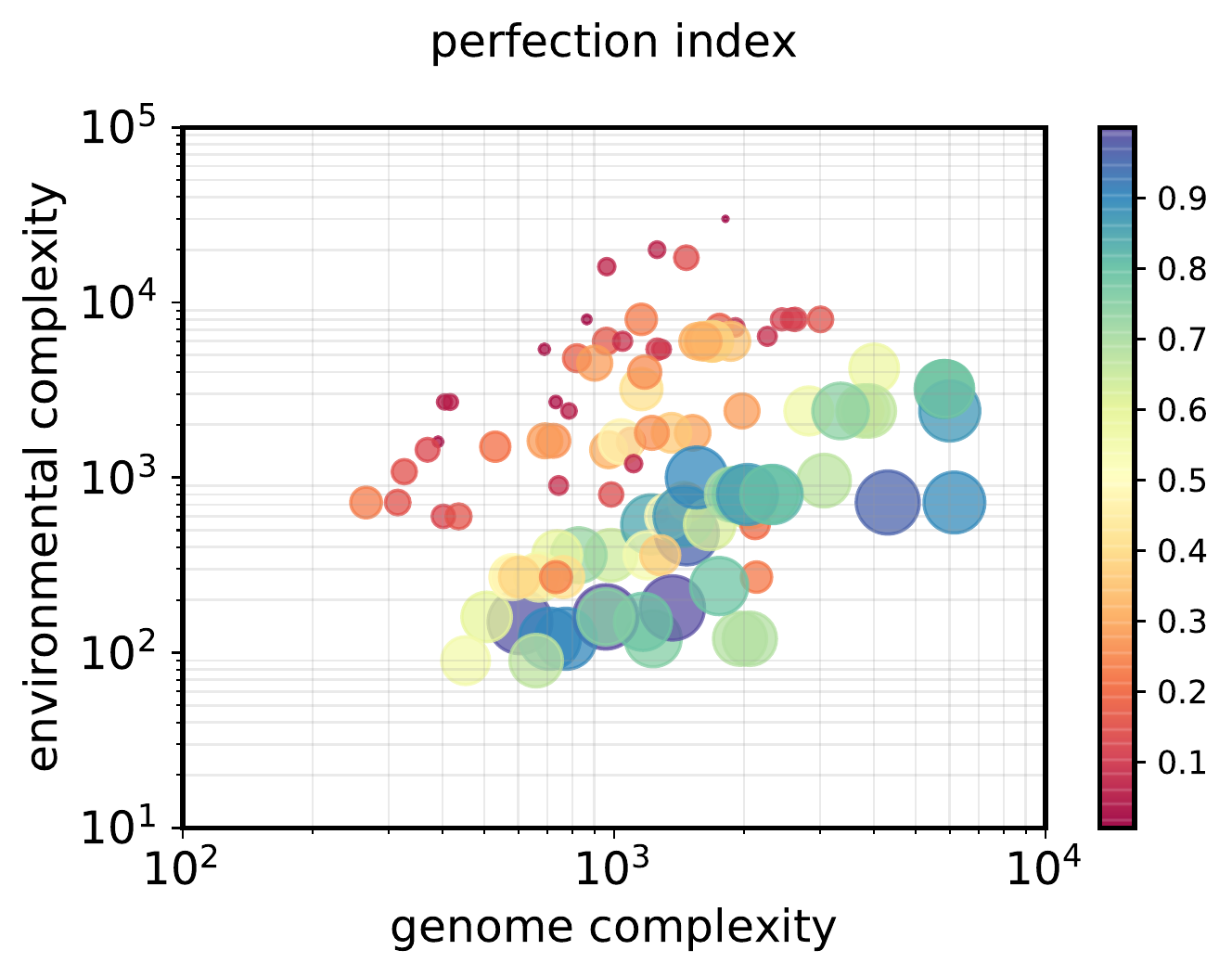}
    \caption{Selection is inefficient for runs with a combination of high environmental complexity and low genome complexity. $X$ axis: genome complexity (number of genes in each organism times the number of types of bases of which those genes are made up, times the average length of the genes at the end of the selection process. $Y$ axis: environmental complexity; number of environmental factors that must match the genotype, multiplied by the number of different environments to be fitted, the number of base types in each sequence, and the length of the environmental strings to be matched. The ‘perfection index’ --fitness at selection plateau divided by maximum possible fitness-- is both proportional to the circle areas and, for enhanced readability, to the colour scale (vertical bar). }
\end{figure}

\begin{figure}[!h]
    \centering
    \includegraphics[width=.65\textwidth]{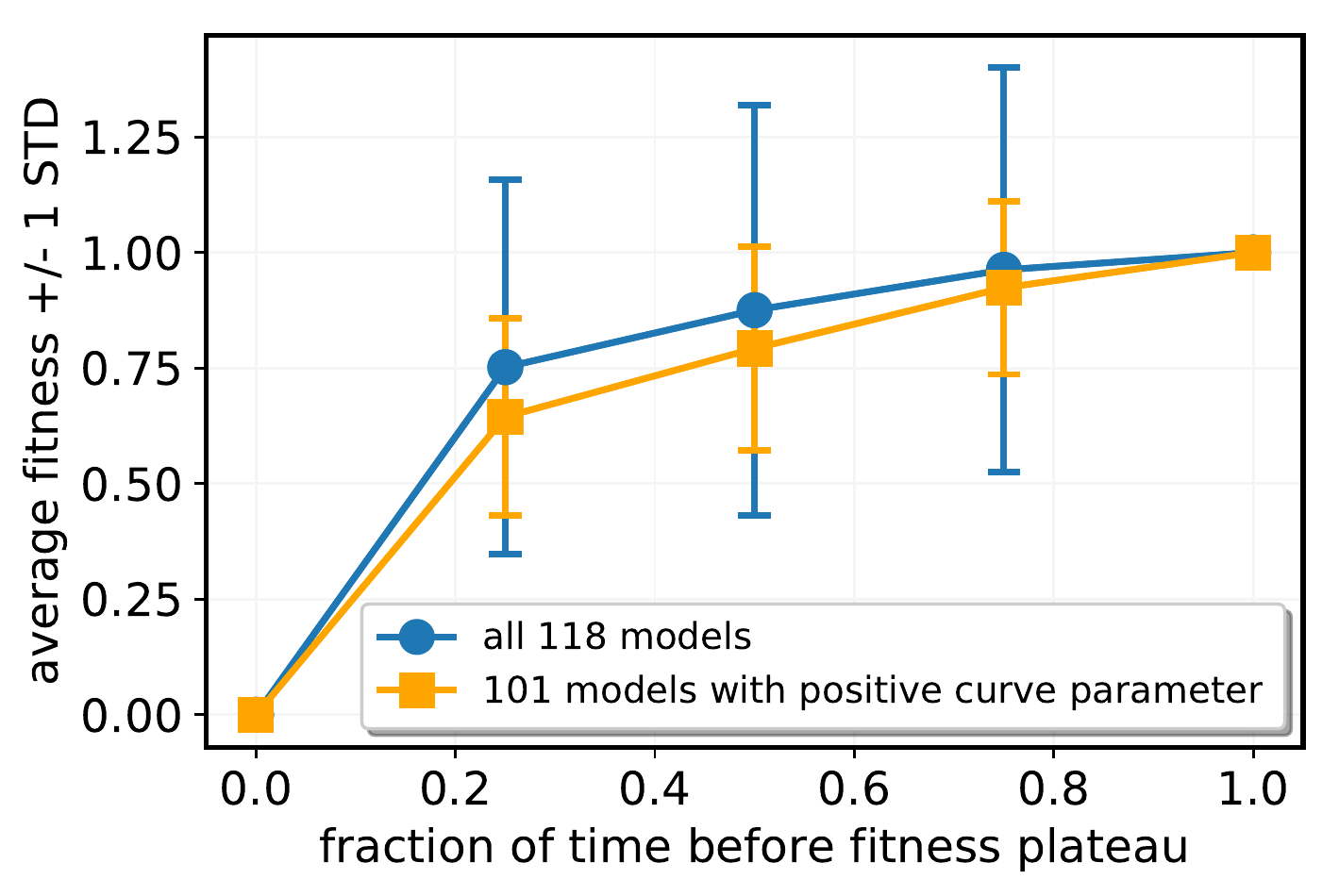}
    \caption{Average fitness of the model runs. Curves are normalized to maximum fitness = 1. The $X$ axis shows the fraction of time until the fitness reaches a stable plateau. The average fitness across all 118 model runs was calculated for time $\in[0,0.25]$, time $\in[ 0.25, 0.5]$ , time $\in[0.5,0.75]$, time $\in[0.75,1]$ and $\textrm{time} >1$ (i.e. on the plateau of fitness). By definition, fitness = 1 when $\textrm{time} >1$. Error bars are standard deviations. Blue curve: All 118 model runs. Orange curve: 101 model runs for which Curve Parameter $C_p>0$. }
\end{figure}

    \subsection{Coding region evolution}

There is a slight bias the mutation mechanism towards gene shrinkage, included because a) this is seen in real mutation rates and b) it protects the model against indefinite expansion of genes through ‘drift’. Despite this, the average length of coding regions tends to increase with model progression (Figure 6). This is explicable as follows. Gene activation depends on matching part of an expressed coding sequence to a regulatory sequence. Thus larger genes mean a greater chance of productive interaction with a regulatory element. The only selective pressure against long genes is the chance that they interact with one of the ‘negative’ environmental elements when they are expressed. As for all runs there are more regulatory elements (20 per gene) than environmental factors (a maximum of 200), this provides a selection pressure towards longer genes.

\begin{figure}
    \centering
    \includegraphics[width=.65\textwidth]{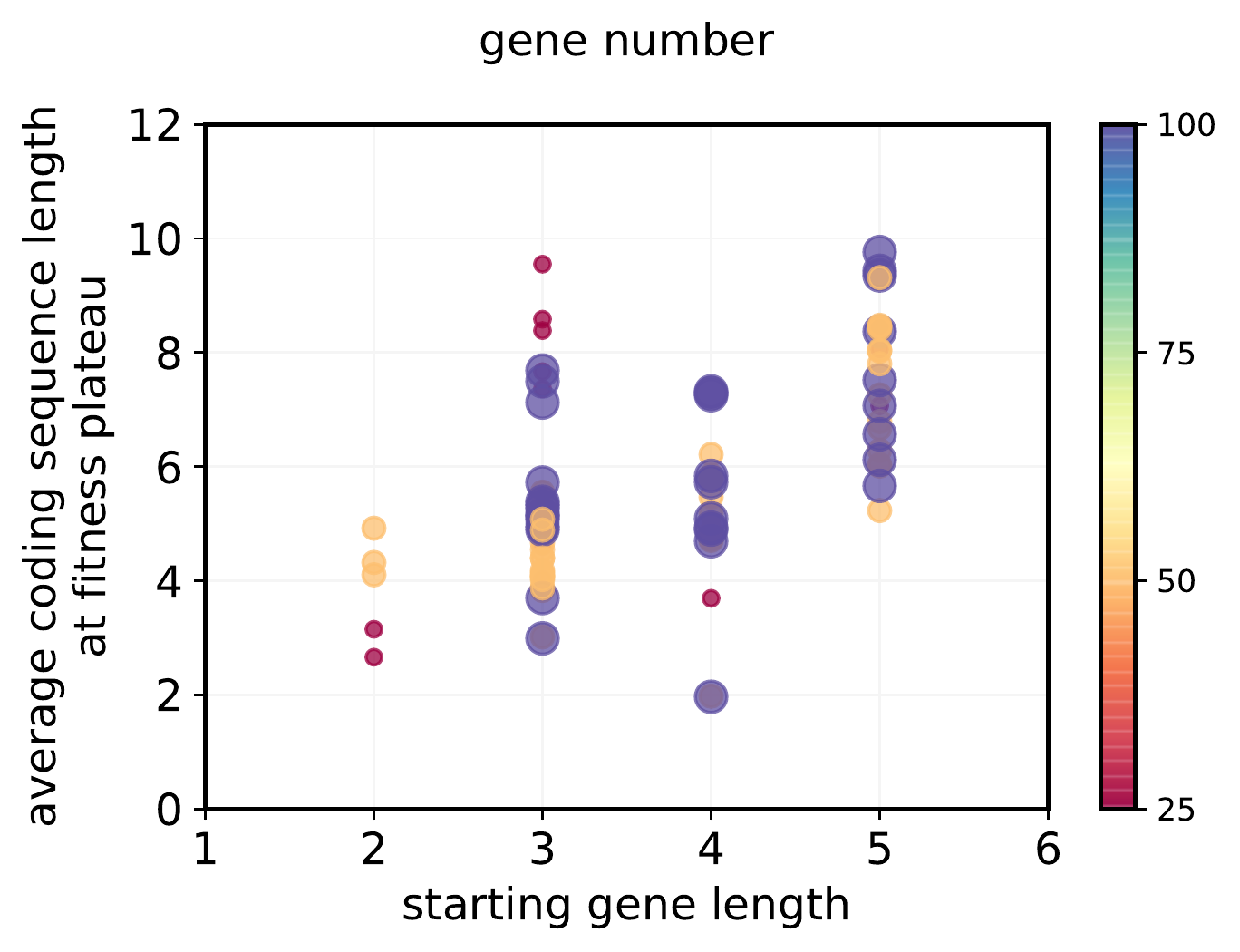}
    \caption{Length of the coding sequences averaged across all genes in a run ($Y$ axis) as a function of the starting length of those sequences ($X$ axis). Circle sizes and color scale show the number of genes in the run, and show no distinct pattern. }
\end{figure}

    \subsection{Default genetic control measures}

The computational effort to exhaustively analyse the entire control network  of up to 100 genes each interacting with up to 20 control elements in each gene in each of 5 organisms in 100 runs or up to $50'000$ time steps each is unrealistic, and so we summarise the overall style of control as follows.\\ 

\noindent
{\bf Average control element length.} A short regulatory element is more likely to match a sequence in the phenotype than a long regulatory element, because a short string is more likely to match a random target by chance than is a long string. (Consider the chance that the strings “A” and “ALPHABET” will match the text in this paper) Thus if negative regulatory elements are on average shorter than positive regulatory elements in a gene, it is likely that the gene is not active. Thus the ratio 

\[
R_a = \textrm{ ratio (average regulatory element length)}
\]
\[
\ =\frac{\sum \textrm{neg. elem. len.} / \sum \textrm{neg. elem. count}}
{\sum \textrm{pos. elem. len.} / \sum \textrm{pos. elem. count}}
\]

\noindent
is a measure of the bias towards inactive genes, i.e. of ‘default off’ regulatory style. (Zero-length elements, i.e. ones which have been deleted, are not counted in the average).\\

\noindent
{\bf Average minimum control element length.} The problem with average regulatory element length as a measure of control logic is that one short regulatory element (likely to be active) can dominate the control of a gene over a number of long regulatory elements (which are in effect ‘junk DNA’, never being active). So the shortest regulatory element is the one most likely to be ‘active’ in a gene. If the shortest positive regulatory element is shorter than the shortest negative regulatory element, then there is a greater chance that the gene will be active; if the shortest negative regulatory element is shorter than the shortest positive element, then the gene is more likely to be inactive. We therefore adopted a measure looking for the shortest regulatory element in a gene. For each gene, the shortest non-zero control sequence is recorded for positive and negative control elements. The average of the length of the shortest regulatory element for all genes in the genotype is reported.  Thus the ratio 

\[
R_m = \textrm{ ratio (minimum regulatory element length)}
\]
\[
\ =\frac{(\sum \textrm{min. neg. elem. len./gene} )/ \sum \textrm{genes}}
{(\sum \textrm{min. pos. elem. len./gene}) / \sum \textrm{genes}}
\]

\noindent
is a measure of the bias towards inactive genes, i.e. of ‘default off’ regulatory style. (Again, zero-length elements, i.e. ones which have been deleted, are not counted in the average) 

Correlations of these two measures to both the inputs and the outputs of model runs are provided in Table 1. We emphasise that this is an initial modelling study, and much more extensive modelling with more efficiently coded models and better hardware will be needed to confirm, expand and dissect these findings. However three patterns are clear from the results here. 

\begin{table}[]
\begin{tabular}{|l|l|l|}
\hline
\textbf{Model parameters}                                               & \multicolumn{2}{l|}{\textbf{Measures of genetic control logic}}                                                                                                             \\ \hline
                                                                        & \begin{tabular}[c]{@{}l@{}}min $-$ve / min $+$ve\\  ($<1 =$ default off)\end{tabular} & \begin{tabular}[c]{@{}l@{}}avg -ve / avg $+$ve\\  ($<1 =$ default off)\end{tabular} \\ \hline
\multicolumn{3}{|c|}{\textbf{Starting parameters}}                                                                                      \\ \hline
Number of environments                                                  & $-$0.02                                                                               & 0.032                                                                               \\ \hline
Number of environmental factors                                         & 0.262 *                                                                               & 0.144                                                                               \\ \hline
\begin{tabular}[c]{@{}l@{}}Environmental complexity\\  (a)\end{tabular} & 0.141                                                                                 & 0.151                                                                               \\ \hline
Length of initial gene                                                  & 0.445 **                                                                              & $-$0.044                                                                            \\ \hline
Number of genes                                                         & $-$0.144                                                                              & $-$0.144                                                                            \\ \hline
Genome complexity at start (b)                                          & 0.297 **                                                                              & 0.096                                                                               \\ \hline
\multicolumn{3}{|c|}{\textbf{Parameters at fitness plateau}}                                                                                                                                                                                          \\ \hline
Adaptation at plateau                                                   & 0.326 **                                                                              & 0.167                                                                               \\ \hline
Fraction of perfection                                                  & 0.117                                                                                 & $-$0.015                                                                            \\ \hline
Average gene length                                                     & 0.683 ****                                                                            & 0.198                                                                               \\ \hline
Genome complexity at end (b)                                            & 0.373 **                                                                              & 0.180                                                                               \\ \hline
regulatory complexity at end (c)                                        & 0.13                                                                                  & 0.184                                                                               \\ \hline
Number of genes expressed                                               & 0.262 *                                                                               & 0.102                                                                               \\ \hline
Fraction of genes expressed                                             & 0.713 ****                                                                            & 0.454 **                                                                            \\ \hline
\end{tabular}
\caption{Correlations with control logic style.}
\end{table}

Correlations between two measures of genetic ‘style’ and some inputs and outputs from models. 

\begin{itemize}
    \item [(a)] Environmental complexity is the [number of characters] $\times$ [number of environments] $\times$ [number of factors per environment];
    \item  [(b)] Genome complexity = [number of characters] $\times$ [number of genes] $\times$ [average length of coding regions];
    \item [(c)] Regulatory complexity = [number of characters] $\times$ [number of genes] $\times$ [average length of regulatory elements].
\end{itemize}
  “Min $–$ve” = average length of the shortest negative regulatory element in each gene, averaged over all genes. “min $+$ve“= average length of the shortest positive regulatory element in each gene, averaged over all genes. “Avg $–$ve” = average length of all non-zero-length negative regulatory elements in the genome.”Avg $+$ve”= average length of all non-zero-length positive regulatory elements in the genome. Significance of the correlations (i.e. chance that the observed correlation is seen in 101 model runs if the measures of control logic are not correlated with the input parameters) $* = p < 0.05, ** = p < 0.01 , *** = p < 0.001, ***\,* = p < 0.0001$.  

Firstly, whether default on or default off genetics evolves is largely independent of how complex the environment is. This was a surprising result, as we expected more complex environments to drive selection for more complex genetic controls, with consequences (positive or negative) for a ‘default off’ control style. From this initial analysis, such effects do not appear to dominate. This may be an artefact of the small sizes of the populations (leading to noise which obscures patterns), the small number of combinations of parameters explored (61 out of 720 possible combinations of the parameters used in these runs), or the small genomes (maximum 100 genes). 

Secondly, genome complexity both at the start and at the end of the models is correlated with ‘default on’ control style. Again, this was a surprise. Our hypothesis is that ‘default off’ genetics allows complex genome evolution. However, our initial hypothesis, that ‘default off’ genetics allows ready gene duplication,  is not captured in this model, where the number of genes is fixed.  

Lastly, ‘default on’ genetics is most strongly correlated with the number of expressed genes. We further dissect this in Figure 7. There is a striking correlation between two measures of ‘default control logic’ and the number of expressed genes. If relatively few genes are expressed in a genome, then ‘default off’ is preferred. If many genes are expressed, ‘default on’ is preferred. Note that the correlation with the number of expressed genes is much weaker (Table 1)– it is the fraction of the genome which is expressed that correlates more strongly with genetic logic than any other parameter. 

\begin{figure}[!h]
    \centering
    \includegraphics[width=.65\textwidth]{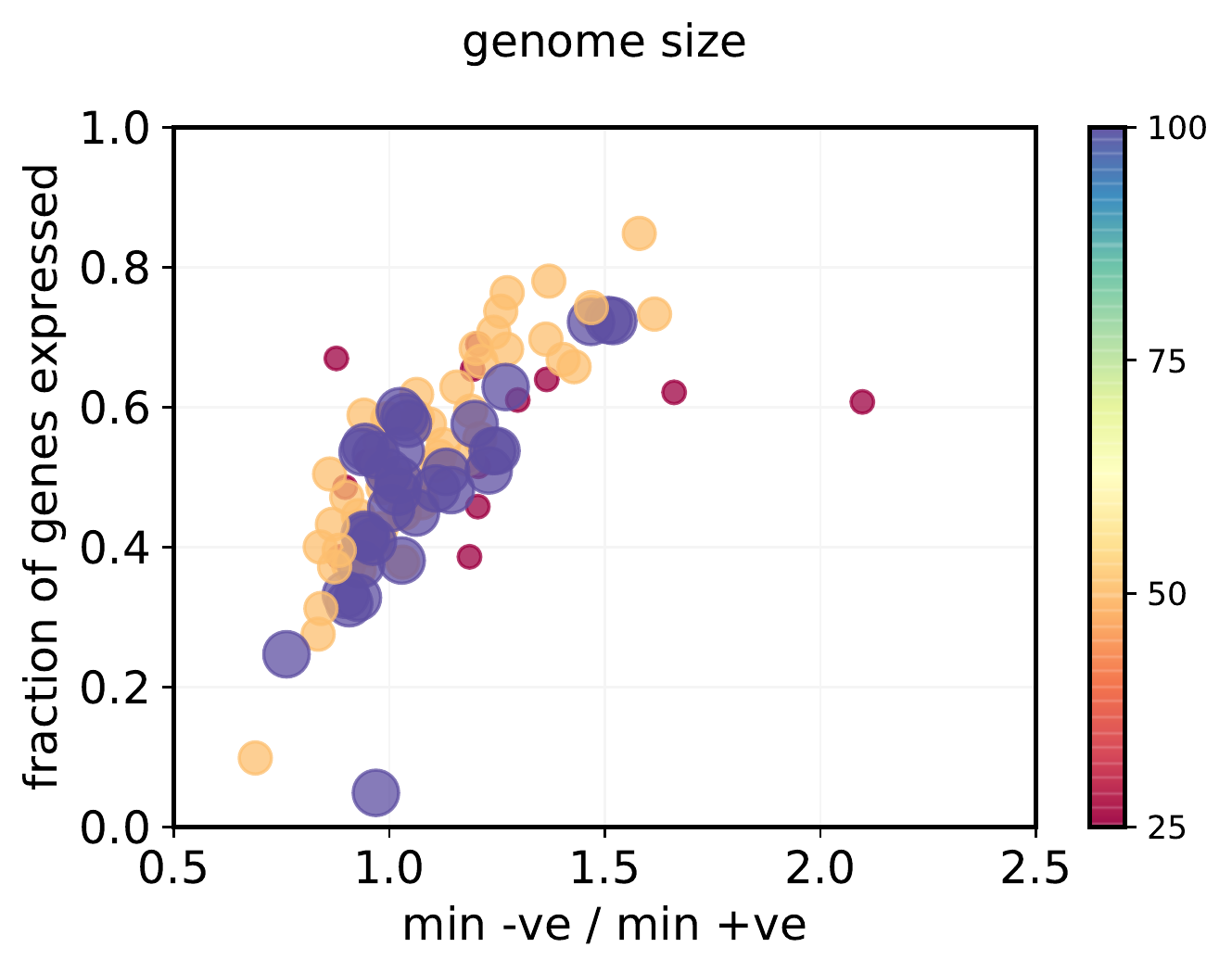}
    \caption{Relationship between the ratio of minimum negative elements to minimum positive elements ($X \textrm{ axis:} <1 = $‘default off’) to the fraction of genes in a genome expressed at fitness plateau ($Y$ axis). Both circle area and color scale are proportional to genome size (25, 50 or 100 genes).  “Min $–$ve” = average length of the shortest negative regulatory element in each gene, averaged over all genes. “min $+$ve“= average length of the shortest positive regulatory element in each gene, averaged over all genes.}
\end{figure}

    \subsection{Reproducibility}
 
No stochastic model will give the same results in different runs, so it is important to show that the variability in output is not so extreme as to render results uninterpretable. The purpose of this modelling was to test the model concept and provide an initial exploration of parameter space: as a result, only a few sets of runs of the model were replicate runs with the same parameters. We chose three runs that gave different control logic outputs in an initial run and re-ran the same parameters with different starting genomes and environments. The results are summarised in Figure 8. This shows that, while results are variable, the genetic control outputs, and specifically the $R_m$ parameter, are consistent within replicates: Replicates of a model run that gave $R_m = 1$ consistently gave $R_m \sim 1$ (``Replicate 1''), Replicates of a run which gave $R_m > 1$ (``Replicate 2'') consistently gave $R_m > 1$, and Replicates of a run that gave  $R_m < 1$ consistently gave $R_m < 1$ (``Replicate 3'')

\begin{figure}
    \centering
    A)\hfill\phantom{0}\\
    \includegraphics[width=.65\textwidth]{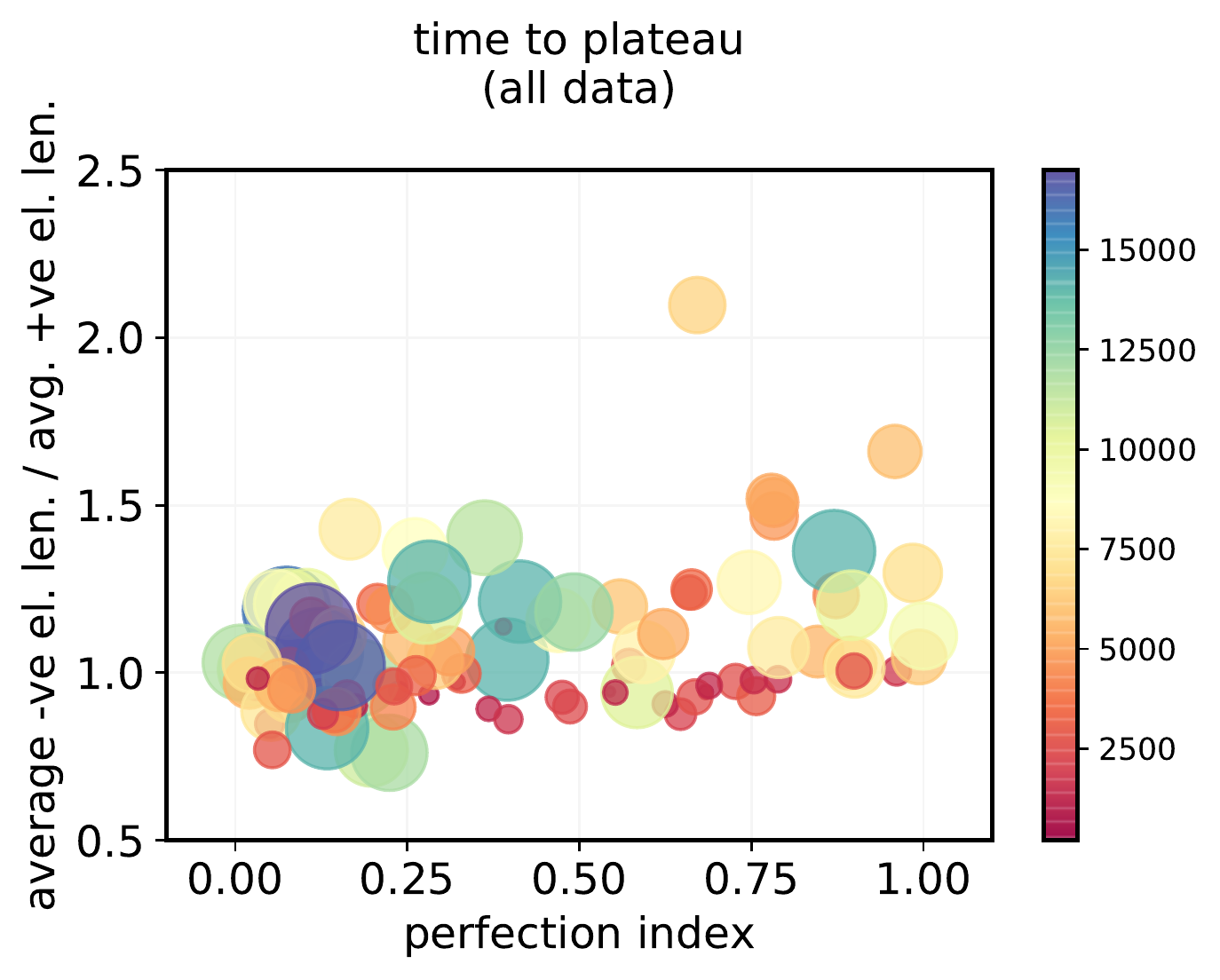}\\
    B)\hfill\phantom{0}\\
    \includegraphics[width=.65\textwidth]{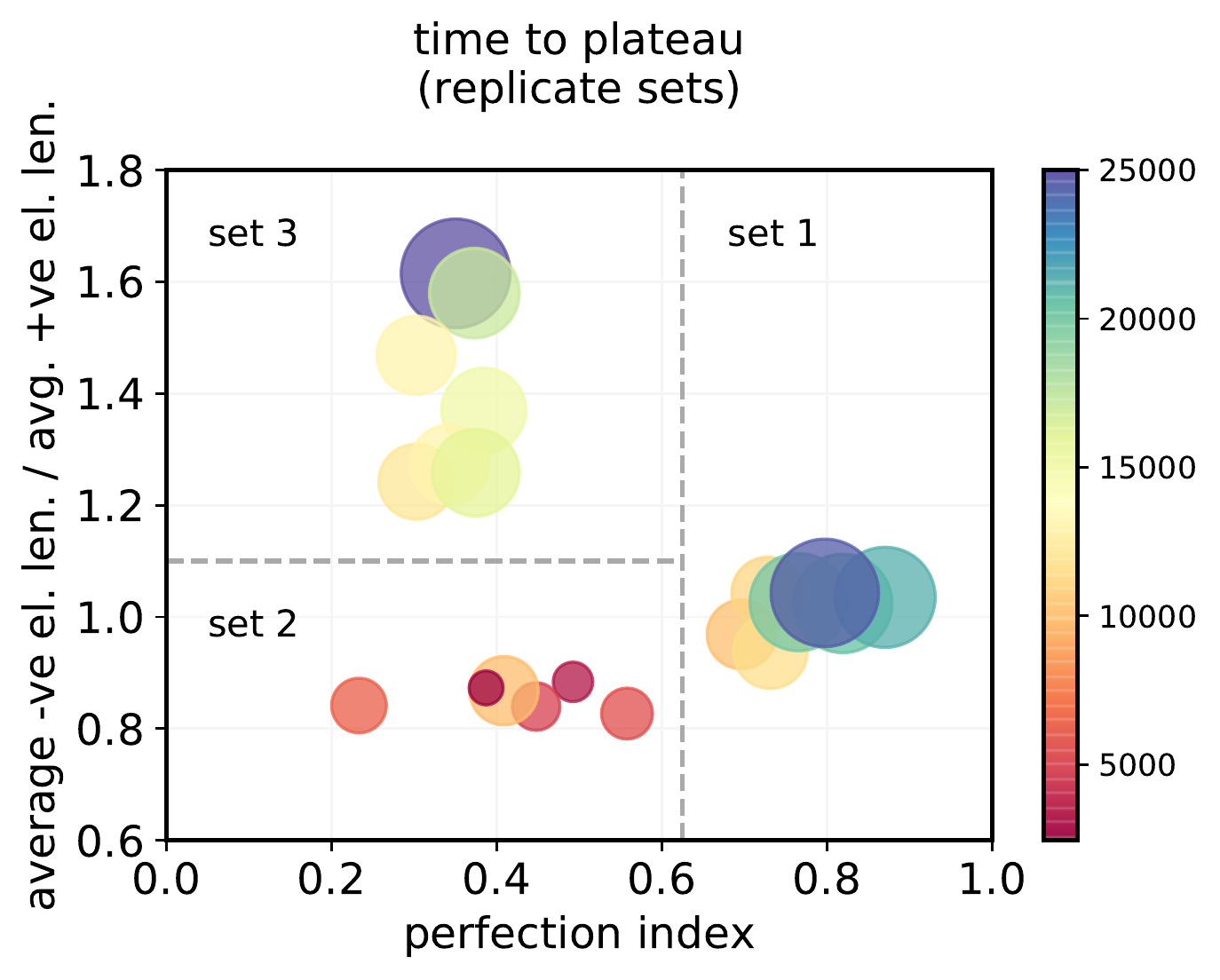}
    \caption{Reproducibility across runs. A) Example outcomes from all runs with diverse starting conditions, and B) From replicate sets of runs started from the same set of parameters. $X$ axis: ‘Perfection index’ (fitness at the fitness plateau as a fraction of the maximum possible fitness with those parameters). $Y$ axis: ratio of the length of the average minimum negative regulatory element length to the average minimum positive regulatory element length. Both circle area and color scale are proportional to the number of steps taken to reach a fitness plateau.  }
    \label{fig:my_label}
\end{figure}

\section{Discussion and Conclusions} 

We have presented a model of the evolution of genome control logic, and an initial analysis of its performance on a small number of test cases. The model performs in a comprehensible way, and evolves fitter organisms. Preliminary statistics suggest that the model performance is stable, i.e. a given set of starting conditions will give a set of outputs more closely related to each other than random, despite the model being a stochastic one.  

We emphasise that this is a preliminary exploration of this model only, and much more needs to be done. However with that caveat, the results show three things of potential interest to the hypothesis that stimulated its creation 

\begin{itemize}
    \item [{\it i}] The genetic logic a population of organisms evolves is not related to the complexity of the environment it finds itself in. This was unexpected
    \item [{\it ii})] The evolved genetic logic is strongly related to the starting and the final, evolved genome complexity. More complex genomes have ‘default on’ logic. This is not predicted by the model, but as the model’s predictions on evolution of genetic logic refers primarily to the acquisition of new genes in the genome, an aspect of evolution not captured here, this does not test the hypothesis. 
    \item [{\it iii})] The strongest correlations with genetic logic are with the fraction of the genome that is expressed.  
\end{itemize}

Point ({\it iii}) above fits with (although is a weak test of) our original hypothesis. It also fills in a significant gap in the hypothesis about why a ‘default off’ logic should be selected. Clearly, an organism cannot evolve ‘default off’ in anticipation of acquiring new genes. However if a specific combination of environmental and genetic features encouraged the development of ‘default off’ genetics, then such an organism would be pre-adapted for genome complexification by gene duplication and divergence. As noted in the introduction, the majority of genes in eukaryotic genomes are not expressed at any one time. Most of them are ‘off’. Our model appears to be evolving a similar expression pattern in some cases, and in those cases the ‘default off’ logic is selected.  

If our results represent the more complex world of real genetics, then we might speculate that organisms living in an environment that occasionally called on a diverse set of genes but most of the time did not require them would feel short-term selective pressure to evolve a ‘default off’ logic. Such an environment could be one in which a heterotroph lived in a community made up of a changing composition of autotrophs, each of which provided a small number of substrates to the heterotroph.  If such a scenario were valid, then we would expect more comprehensive modelling to reveal influences of environmental change on both expressed gene numbers and default logic. Such work is being actively pursued.  

\section*{References} 

\begin{enumerate}

\item
	Knoll AH. The Multiple Origins of Complex Multicellularity. Annual Review of Earth and Planetary Sciences. 2011;39(1):217-39. 

\item
	Parfrey LW, Lahr DJG. Multicellularity arose several times in the evolution of eukaryotes (Response to DOI 10.1002/bies.201100187). BioEssays. 2013;35(4):339-47. 

\item
	Rokas A. The Origins of Multicellularity and the Early History of the Genetic Toolkit For Animal Development. Annual Review of Genetics. 2008;42(1):235-51. 

\item
	Dagan T, Roettger M, Stucken K, Landan G, Koch R, Major P, et al. Genomes of Stigonematalean Cyanobacteria (Subsection V) and the Evolution of Oxygenic Photosynthesis from Prokaryotes to Plastids. Genome biology and evolution. 2013;5(1):31-44. 

\item
 	Chang Y-j, Land M, Hauser L, Chertkov O, Glavina Del Rio T, Nolan M, et al. Non-contiguous finished genome sequence and contextual data of the filamentous soil bacterium Ktedonobacter racemifer type strain (SOSP1-21T). Standards in Genomic Sciences. 2011;5(1):97-111. 

\item
	Schneiker S, Perlova O, Kaiser O, Gerth K, Alici A, Altmeyer MO, et al. Complete genome sequence of the myxobacterium Sorangium cellulosum. Nat Biotech. 2007;25(11):1281-9. 

\item
	Kolinko S, Richter M, Glöckner F-O, Brachmann A, Schüler D. Single-cell genomics of uncultivated deep-branching magnetotactic bacteria reveals a conserved set of magnetosome genes. Environmental Microbiology. 2016;18(1):21-37. 

\item
	Xu J, Saunders CW, Hu P, Grant RA, Boekhout T, Kuramae EE, et al. Dandruff-associated Malassezia genomes reveal convergent and divergent virulence traits shared with plant and human fungal pathogens. Proceedings of the National Academy of Sciences. 2007;104(47):18730-5. 

\item
	Anantharaman V, Iyer LM, Aravind L. Comparative Genomics of Protists: New Insights into the Evolution of Eukaryotic Signal Transduction and Gene Regulation. Annual review of microbiology. 2007;61(1): 453-75. 

\item
	Adams MD, Celniker SE, Holt RA, Evans CA, Gocayne JD, Amanatides PG, et al. The Genome Sequence of Drosophila melanogaster. Science. 2000;287(5461):2185-95. 

\item
	Borukhov S, Lee J, Laptenko O. Bacterial transcription elongation factors: new insights into molecular mechanism of action. Molecular Microbiology. 2005;55(5):1315-24. 

\item
	Lander ES, Linton LM, Birren B, Nusbaum C, Zody MC, Baldwin J, et al. Initial sequencing and analysis of the human genome. Nature. 2001;409:860. 

\item
	Ponting CP, Hardison RC. What fraction of the human genome is functional? Genome Research. 2011;21(11):1769-76. 

\item
	Ferdows MS, Barbour AG. Megabase-sized linear DNA in the bacterium Borrelia burgdorferi, the Lyme disease agent. Proceedings of the National Academy of Sciences. 1989;86(15):5969-73. 

\item
	Hinnebusch J, Tilly K. Linear plasmids and chromosomes in bacteria. Molecular Microbiology. 1993;10(5):917-22. 

\item
	Sherman L, Min H, Toepel J, Pakrasi H. Better Living Through Cyanothece – Unicellular Diazotrophic Cyanobacteria with Highly Versatile Metabolic Systems. In: Hallenbeck PC, editor. Recent Advances in Phototrophic Prokaryotes. Advances in Experimental Medicine and Biology. 675: Springer New York; 2010. p. 275-90. 

\item
	Kube M, Schneider B, Kuhl H, Dandekar T, Heitmann K, Migdoll A, et al. The linear chromosome of the plant-pathogenic mycoplasma 'Candidatus Phytoplasma mali'. BMC Genomics. 2008;9(1):306. 

\item
	Kulp A, Kuehn MJ. Biological Functions and Biogenesis of Secreted Bacterial Outer Membrane Vesicles. Annual review of microbiology. 2010;64:163-84. 

\item
	Stolz JF. Bacterial Intracellular Membranes.  eLS: John Wiley \& Sons, Ltd; 2001. 

\item
	Hanson RS, Hanson TE. Methanotrophic bacteria. Microbiological Reviews. 1996;60(2):439-71. 

\item
	Prust C, Hoffmeister M, Liesegang H, Wiezer A, Fricke WF, Ehrenreich A, et al. Complete genome sequence of the acetic acid bacterium Gluconobacter oxydans. Nat Biotech. 2005;23(2):195-200. 

\item
	Fuerst JA. Intracellular compartmentalization in Planctomycetes. Ann Rev Microbiology. 2005;59:299 - 328. 

\item
	van Niftrik LA, Fuerst JA, Damsté JSS, Kuenen JG, Jetten MSM, Strous M. The anammoxosome: an intracytoplasmic compartment in anammox bacteria. FEMS Microbiology Letters. 2004;233(1):7-13.

\item
	van Niftrik L, Geerts WJC, van Donselaar EG, Humbel BM, Yakushevska A, Verkleij AJ, et al. Combined structural and chemical analysis of the anammoxosome: A membrane-bounded intracytoplasmic compartment in anammox bacteria. Journal of Structural Biology. 2008;161(3):401-10. 

\item
	Fuerst JA, Webb RI, Garson MJ, Hardy L, Reiswig HM. Membrane-bounded nucleoids in microbial symbionts of marine sponges. FEMS Microbiology Letters. 1998;166(1):29-34.

\item
	Schorn S, Salman-Carvalho V, Littmann S, Ionescu D, Grossart H-P, Cypionka H. Cell Architecture of the Giant Sulfur Bacterium Achromatium oxaliferum: Extra-cytoplasmic Localization of Calcium Carbonate Bodies. FEMS Microbiology Ecology. 2019;96(2). 

\item
	Keren R, Mayzel B, Lavy A, Polishchuk I, Levy D, Fakra SC, et al. Sponge-associated bacteria mineralize arsenic and barium on intracellular vesicles. Nature Communications. 2017;8(1):14393. 

\item
	Kunkel DD. Thylakoid centers: Structures associated with the cyano- bacterial photosynthetic membrane system. Archives of Microbiology. 1982;133(2):97-9. 

\item
	Jékely G. Origin of eukaryotic endomembranes. In: Jékely G, editor. Eukaryotic Membranes and Cytoskeleton: Origins and Evolution. Berlin: Springer; 2007. p. 38 - 51. 

\item
	Roeben A, Kofler C, Nagy I, Nickell S, Ulrich Hartl F, Bracher A. Crystal Structure of an Archaeal Actin Homolog. Journal of Molecular Biology. 2006;358(1):145-56. 

\item
	Vollmer W. The prokaryotic cytoskeleton: a putative target for inhibitors and antibiotics? Applied Microbiology and Biotechnology. 2006;73(1):37-47. 

\item
	van den Ent F, Amos LA, Lowe J. Prokaryotic origin of the actin cytoskeleton. Nature. 2001;413(6851):39-44. 

\item
	Erickson HP. FtsZ, a tubulin homologue in prokaryote cell division. Trends in Cell Biology. 1997;7(9):362-7. 

\item
	Ausmees N, Kuhn JR, Jacobs-Wagner C. The Bacterial Cytoskeleton: An Intermediate Filament-Like Function in Cell Shape. Cell. 2003;115(6):705-13. 

\item
	Williams TA, Foster PG, Cox CJ, Embley TM. An archeal origin of eukaryotes supports only two primary domains of life. Nature. 2014;504:231 - 6. 

\item
	Montgomery WL, Pollak PE. Epulopiscium fishelsoni N. G., N. Sp., a Protist of Uncertain Taxonomic Affinities from the Gut of an Herbivorous Reef Fish. Eukaryotic microbiology. 1988;35(4):565 - 9. 

\item
	Angert ER, Clements KD, Pace NR. The largest bacterium. Nature. 1993;362(6417):239-41. 

\item
	de Duve C. The origin of eukaryotes: a reappraisal. Nature Reviews Genetics. 2007;8:395 - 403. 

\item
	Pittis AA, Gabaldón T. Late acquisition of mitochondria by a host with chimaeric prokaryotic ancestry. Nature. 2016;531(7592):101 - 4. 

\item
	Thao ML, Gullan PJ, Baumann P. Secondary ($\gamma$-Proteobacteria) Endosymbionts Infect the Primary ($\beta$-Proteobacteria) Endosymbionts of Mealybugs Multiple Times and Coevolve with Their Hosts. Applied and Environmental Microbiology. 2002;68(7):3190-7. 

\item
	Wujek DE. Intracellular bacteria in the blue-green alga Pleurocapsa minor. Transactions of the American Microscopic Society. 1979;98(1): 143 - 5. 

\item
	Larkin JM, Henk MC. Filamentous sulfide-oxidizing bacteria at hydrocarbon seeps of the Gulf of Mexico. Microscopy research and technique. 1996;33(1):23-31. 

\item
	von Dohlen CD, Kohler S, Alsop ST, McManus WR. Mealybug [beta]-proteobacterial endosymbionts contain [gamma]-proteobacterial symbionts. Nature. 2001;412(6845):433-6. 

\item
	Sassera D, Beninati T, Bandi C, Bouman EAP, Sacchi L, Fabbi M, et al. ‘Candidatus Midichloria mitochondrii’, an endosymbiont of the tick Ixodes ricinus with a unique intramitochondrial lifestyle. International Journal of Systematic and Evolutionary Microbiology. 2006;56(11): 2535-40. 

\item
	Lane N. Energetics and genetics across the prokaryote-eukaryote divide. Biol Direct. 2011;6:35. 

\item
	Lane N, Martin W. The energetics of genome complexity. Nature. 2010;467(7318):929-34. 

\item
	Lane N, Martin WF. Eukaryotes really are special, and mitochondria are why. Proceedings of the National Academy of Sciences. 2015; 112(35):E4823. 

\item
	Tijhuis L, Van Loosdrecht MCM, Heijnen JJ. A thermodynamically based correlation for maintenance gibbs energy requirements in aerobic and anaerobic chemotrophic growth. Biotechnology and Bioengineering. 1993;42(4):509-19. 

\item
	Hoehler T. Biological energy requirements as quantitative boundary conditions for life in the subsurface. Geobiology. 2004;2(4):205-15. 

\item
	Hoehler TM, Amend J, Shock EL. A “Follow the Energy” Approach for Astrobiology. Astrobiology. 2007;7(6):819-23. 

\item
	Bains W, Schulze-Makuch D. Mechanisms of Evolutionary Innovation Point to Genetic Control Logic as the Key Difference Between Prokaryotes and Eukaryotes. Journal of Molecular Evolution. 2015:1-20 (DOI 10.1007/s00239-015-9688-6). 

\item
	Bains W, Schulze-Makuch D. The Cosmic Zoo: The (Near) Inevitability of the Evolution of Complex, Macroscopic Life. Life. 2016;6(3):25. 

\item
	Wagner FR, Dienemann C, Wang H, Stützer A, Tegunov D, Urlaub H, et al. Structure of SWI/SNF chromatin remodeller RSC bound to a nucleosome. Nature. 2020;579(7799):448-51. 

\item
	Ye Y, Wu H, Chen K, Clapier CR, Verma N, Zhang W, et al. Structure of the RSC complex bound to the nucleosome. Science. 2019;366(6467): 838-43. 

\item
	Cairns BR. Chromatin remodeling machines: similar motors, ulterior motives. Trends in Biochemical Sciences. 1998;23(1):20-5. 

\item
	Kireeva ML, Walter W, Tchernajenko V, Bondarenko V, Kashlev M, Studitsky VM. Nucleosome Remodeling Induced by RNA Polymerase II: Loss of the H2A/H2B Dimer during Transcription. Molecular Cell. 2002;9(3):541-52. 

\item
	Mizuguchi G, Shen X, Landry J, Wu W-H, Sen S, Wu C. ATP-Driven Exchange of Histone H2AZ Variant Catalyzed by SWR1 Chromatin Remodeling Complex. Science. 2004;303(5656):343-8. 

\item
	Shen X, Mizuguchi G, Hamiche A, Wu C. A chromatin remodelling complex involved in transcription and DNA processing. Nature. 2000;406(6795):541-4. 

\item
	Olave IA, Peck-Peterson SI, Crabtree GR. Nuclear actin and actin-related proteins in chromatin remodelling. Annual Review of Biochemistry. 2002;71:755 - 81. 

\item
	Grayling RA, Sandman K, Reeve JN. Histones and chromatin structure in hyperthermophilic Archaea. FEMS Microbiology Reviews. 1996;18(2-3):203-13. 

\item
	Sandman K, Reeve JN. Archaeal histones and the origin of the histone fold. Current Opinion in Microbiology. 2006;9(5):520-5. 

\item
	Saecker RM, Record MT, deHaseth PL. Mechanism of Bacterial Transcription Initiation: RNA Polymerase - Promoter Binding, Isomerization to Initiation-Competent Open Complexes, and Initiation of RNA Synthesis. Journal of Molecular Biology. 2011;412(5):754-71.

\item
	Browning DF, Busby SJW. The regulation of bacterial transcription initiation. Nature Reviews Microbiology. 2004;2(1):57-65. 

\item
	Reeve JN. Archaeal chromatin and transcription. Molecular Microbiology. 2003;48(3):587-98. 

\item
	Geiduschek EP, Ouhammouch M. Archaeal transcription and its regulators. Molecular Microbiology. 2005;56(6):1397-407. 

\item
	Tsukiyama T, Wu C. Chromatin remodeling and transcription. Current Opinion in Genetics \& Development. 1997;7(2):182-91. 

\item
	McGee MD, Borstein SR, Meier JI, Marques DA, Mwaiko S, Taabu A, et al. The ecological and genomic basis of explosive adaptive radiation. Nature. 2020;586(7827):75-9. 

\item
	Crozat E, Philippe N, Lenski RE, Geiselmann J, Schneider D. Long-term experimental evolution in Escherichia coli. XII. DNA topology as a key target of selection. Genetics. 2005;169(2):523-32. 

\item
	Philippe N, Crozat E, Lenski RE, Schneider D. Evolution of global regulatory networks during a long-term experiment with Escherichia coli. Bioessays. 2007;29(9):846-60. 

\item
	Lamrabet O, Plumbridge J, Martin M, Lenski RE, Schneider D, Hindré T. Plasticity of promoter-core sequences allows bacteria to compensate for the loss of a key global regulatory gene. Molecular biology and evolution. 2019;36(6):1121-33. 

\end{enumerate}

\end{document}